\documentclass[fleqn,usenatbib]{mnras}

\usepackage{newtxtext,newtxmath}

\usepackage[T1]{fontenc}

\DeclareRobustCommand{\VAN}[3]{#2}
\let\VANthebibliography\thebibliography
\def\thebibliography{\DeclareRobustCommand{\VAN}[3]{##3}\VANthebibliography}

\usepackage{graphicx}	
\usepackage{amsmath}	

\title[Partially ergodic SFR deviations]{Examining partial ergodicity as a predictor of star formation departures from the galactic main sequence in isolated galaxies}

\author[F. M. Smith \& R. J. Thacker]{
Fraser M. Smith,$^{1}$\thanks{E-mail: fraser.smith@smu.ca}
\& Robert J. Thacker$^{1}$
\\
$^{1}$Department of Astronomy and Physics and Institute for Computational Astrophysics, Saint Mary's University, Halifax, NS B3H 3C3, Canada\\
}

\date{Accepted XXX. Received YYY; in original form ZZZ}

\pubyear{2024}

\begin{document}
\label{firstpage}
\pagerange{\pageref{firstpage}--\pageref{lastpage}}
\maketitle

\begin{abstract}
Lacking the ability to follow individual galaxies on cosmological timescales, our understanding of individual galaxy evolution is broadly inferred from population trends and behaviours. In its most prohibitive form, this approach assumes that galactic star formation properties exhibit ergodicity, so that individual galaxy evolution can be statistically inferred via ensemble behaviours. The validity of this assumption is tested through the use of observationally motivated simulations of isolated galaxies. The suite of simulated galaxies is statistically constructed to match observed galaxy properties by using kernel density estimation to create structural parameter distributions, augmented by theoretical relationships where necessary. We also test the impact of different physical processes, such as stellar winds or the presence of halo substructure on the star formation behaviour. We consider the subtleties involved in constraining ergodic properties, such as the distinction between stationarity imposed by stellar wind feedback and truly ergodic behaviour. However, without sufficient variability in star formation properties, individual galaxies are unable to explore the full parameter space. While, as expected, full ergodicity appears to be ruled out, we find reasonable evidence for partial ergodicity, where averaging over mass-selected subsets of galaxies more broadly resembles time averages, where the average largest deviation across physical scenarios is 0.20 dex. As far as we are aware, this the first time partial ergodicity has been considered in an astronomical context, and provides a promising statistical concept. Despite morphological changes introduced by close encounters with dark matter substructure, subhaloes are not found to significantly increase deviations from ergodic assumptions.
\end{abstract}




\section{Introduction}

Inferring the evolution of individual systems from the statistical properties of ensembles, and vice versa, is fundamentally challenging. At one extreme, the presence of confounds can produce ensemble trends that are not reproduced within individual or subset behaviours, commonly known as Simpson's Paradox \citep{simpson1951}. A contrasting situation is ergodic systems, wherein the population average of a given parameter is equivalent to a time average, so that individual behaviour can be inferred from population statistics. This is obviously an attractive ansatz in astronomy since predicting evolution on the basis of population statistics would be a route to inferring evolution of systems on long timescales. For example, a common and powerful statistical method in astronomical research is the use of power spectra \citep[][]{mondal2022}. Inferring distributions for individual systems from power spectra constructed based on population statistics implicitly assumes ergodicity (and thus stationarity), particularly when considering galaxy evolution \citep[][]{wang2020b} and Cosmic Microwave Background analyses \citep[][]{marinucci2010}. In the case of spherically averaged power spectra analyses of 21 cm radio emission \citep[][]{mertens2020}, ergodicity is implicitly assumed, which is important to consider given how the universe can evolve after the Epoch of Reionization \citep{mondal2022}. The ergodic hypothesis is also used in studies of the variability of light flares \citep[][]{chang2015}. The distinction should be made, however, that ergodic properties can be assumed for time-averages of individuals (such as when considering galaxy evolution) or in terms of spatial averages (such as in Cosmic Microwave Background analyses) to compare to the ensemble-average.

The star formation rates (SFRs) and equivalently star formation histories (SFHs) of galaxies are determined by a complex combination of various processes \citep[][]{schaye2010}. SFHs can be influenced by, for example, interactions with external factors \citep[such as ram pressure stripping, e.g.][]{pogg2017}, internal feedback processes \citep[both stellar and from active galactic nuclei, e.g.,][]{springel2005agn}, and accretion events \citep[][]{behroozi2013}. Disentangling the contributions of each of these individual processes is challenging since we are typically limited to the total SFH, despite the processes operating on varying timescales \citep{iyer2020}. Overall, while the understanding of the processes that impact star formation in galaxies has greatly improved in the past few years \citep{gir2020}, this knowledge is still coalescing into a detailed understanding of star formation variability and the way physical phenomena contribute to it, especially in the population context.

The primary mechanism for galaxies to grow is via accretion processes. Galaxy growth via accretion is broad area of research \citep[e.g.][]{rodriguez2016,sawicki2020,graham2023}, which includes both accretion from the surrounding environment (the Circumgalactic Medium, CGM) and more drastic accretion from galaxy-galaxy interactions (e.g. the merging of galaxies). The CGM contains an ample supply of gas \citep[e.g.][]{tumlinson2017,faucher2023} that provides crucial fuel for galactic star formation. Galactic outflows and inflows comprise this complex, multiphase gas making its contribution to galaxy growth essential to consider \citep[][]{arm2017}. While the exact mass of the CGM is difficult to constrain \citep[][]{bregman2007,fang2020}, observational and theoretical evidence suggest the existence of cold, warm, and hot gas \citep[][]{ji2020}. Accretion from the CGM is considered to be either in 'cold' or 'hot' mode, dependent on whether shocked infalling gas has sufficient time to cool \citep{faucher2023}. Hence, the inflowing gas either drops below shock temperatures or remains on the order of the virial temperature of the host halo \citep[][]{birnboim2007}. Given that cold gas flows can deliver material immediately available for star formation directly to the star-forming disc \citep[][]{dekel2009cold,shen2013}, the importance of cooling flows in galaxy evolution is self-evident. 
Mergers between galaxies are also significant accretion processes that drive morphological changes and growth, with the relative frequency and strength of mergers changing with cosmic time \citep[e.g.][]{xu2012,whitney2021}. This also results in the specific star formation rate (sSFR, and equivalently the specific star formation history, sSFH) evolution as a function of mass growth in galaxies changing with time \citep[][]{tasca2015}. While major mergers can produce more profound impacts on the evolution of a given galaxy such as disc destruction, the contribution of minor mergers over time cannot be overlooked \citep[][]{kaviraj2014}, especially in spiral galaxies where disc heating can occur \citep[e.g.][]{read2008,matteo2019}. disc heating due to interactions with subhaloes or 10:1 minor mergers has been found to have a modest impact on galactic SFRs \citep[][]{moster2010disk}. At any given epoch the difference is almost always within a factor of 2, and is usually considerably smaller. The strongest, although perhaps not the most common, source of disc heating arises due to close interactions of the disc with subhaloes with log$_{10}$(M/$M_{\odot}$) $\gtrsim$ 10 \citep{grand2016}. Thus, the relative impact of dark matter substructure on the SFHs of disc systems is important to quantify and constrain in order to better understand SFR variability. 

A motivating paper for this work is \cite{wang2020b}, who use the assumption of ergodicity in order to constrain a temporal power spectrum of the sSFR of star-forming galaxies. Constraining a temporal power spectrum is strongly predictive, as it provides insight into how different timescale processes impact vertical deviations from the star-forming main sequence (SFMS) \citep{caplar2019}. Thus, studying how individual galaxies deviate from the population average/median behaviour provides crucial information that can, given the assumptions, be used to infer how SFHs evolve. 

Discussion of ergodicity in the context of star-forming galaxies inevitably requires the comparison of individual galaxies from the well-known SFMS, which is a tight (0.2 - 0.3 dex) relation between SFR and stellar mass content observed in star-forming galaxies \citep[e.g.][]{brinchmann2000,noeske2007,scholz2023}. The SFMS has been found to evolve over cosmic time with nearly constant scatter \citep[][]{speagle2014}, making it invaluable for inferring galaxy evolution across a vast array of epochs. 

Naturally, to test the assumption of ergodicity, we must utilize galaxy simulations since ergodicity in this specific context involves relating an individual galaxy's time evolution to the average of the galaxy sample at any given epoch, despite the observational motivation of the problem. Observationally we are limited to samples of different galaxies at various stages in their evolution, and while in fundamental sense a true assessment of the ergodicity of star formation properties is not possible within human timeframes, in practice inference via SFHs is a potential way of exploring the concept \citep{wang2020b}. However, such an approach has limitations. For example, there is likely intrinsic, non-ergodic, scatter in star formation behaviours that goes beyond those expected on ergodicity alone, environment being a potential example. Equally importantly, when inferring SFHs through spectral energy distribution (SED) fitting \citep{leja2019}, there are uncertainties and possible bias \citep[][]{haskell2024} in these observationally determined SFHs that further complicates statistical study. Nonetheless, it is possible to include non-ergodicity by modelling it as low frequency contributions to the variance of star formation properties. Under this ansatz, observationally derived SFHs on different timescales, such as 5 Myr and 800 Myr determined via equivalent widths (EWs) of H$\alpha$ and H$\delta$, allow constraints to be placed on temporal power spectral densities describing star formation properties, moreover ratios of these parameters should be less sensitive to non-ergodic behaviours. Observations are thus an essential part of exploring ergodic properties and can be extended out across different cosmological epochs. A system that does not satisfy ergodicity is said to break ergodicity; the breaking is \textit{strong} if there are physical mechanisms/boundaries in place that prevent individuals from exploring the full parameter space \citep[][]{spiechow2016}, whereas \textit{weak} breaking simply occurs when ergodicity is not satisfied. Additionally, even if a population of objects is not truly ergodic, it is possible that population statistics may provide bounds on individual behaviour.

The aim of this study is to test the assumption of ergodicity of sSFR deviations from the SFMS. In addition, we test how different physical scenarios impact any ergodicity present, relative to each other process. Ergodicity is a powerful statistical tool if it can be satisfied, since it allows individual behaviours to be inferred from readily available population statistics. Although star formation processes are unlikely to produce ergodic properties in the SFRs of galaxies, it is nonetheless important to constrain how various physical mechanisms affect ergodicity and to quantify deviations from ergodicity. Broadly, processes that increase variations in deviations might be expected to make ergodicity more common and vice versa. This is relevant in the discussion of the change in SFR behaviours of galaxies as they grow beyond a given size \citep[][]{hop2023} and is thus of considerable interest in studies of high redshift populations \citep[][]{adams2023}.

This paper is structured as follows; in Section~\ref{Methods}, the construction of the simulated galaxy sample is described. In Section~\ref{Results}, our findings are summarized and discussed. Finally, we present our main conclusions and future projects to expand upon these findings in Section~\ref{Conclusions}. 

\section{Methods} \label{Methods}

\subsection{Galaxy Models}

Galaxy models were initialized by generating the basic galactic structure using the MakeNewDisk package provided by Dr. Volker Springel. MakeNewDisk produces disc galaxies with a Hernquist \citep{hernprof1990} dark matter halo, a Hernquist stellar bulge, and an exponential stellar disc with associated gas component similar to the model described in \cite{springel2005disk}. 

Simulations are run using the hydrodynamical simulation code GIZMO\footnote{A public version of the code is available at \url{http://www.tapir.caltech.edu/~phopkins/Site/GIZMO.html}} with a meshless finite-mass hydro solver \citep{hop2015}. Cooling physics is implemented using the methods described in \cite{hop2014,hopkins2018} with an external UV background model \citep{treecool2020}. That is, gas cools according to a cooling curve from 10 - 10$^{10}$ K that considers molecular cooling (at low temperatures) and metal-line cooling (at high temperatures) from 11 species \citep{hop2014}. Star formation, with associated stellar wind feedback, is modeled using the effective equation of state subgrid physics module presented in \cite{springel2003}, which is briefly summarized here; gas particles are treated as two-phase (cold and hot) media to represent the interstellar medium, motivated by \cite{mckee1977}. Gas mass from the cold phase is converted into stellar mass in a probabilistic manner once the set critical density is achieved. The interplay between the cold and hot phases is regulated by energy injected from supernovae feedback immediately as stars form, which evaporates cold mass into hot mass. The parameters for the effective equation of state follow that of \cite{springel2003}, which are motivated by the Schmidt-Kennicutt law \citep{schmidt1959,kennicutt1998schmidt,springel2003}. The wind model is that of \cite{springel2003}, where winds are initially decoupled. Parameter values for the model are a mass-loading factor ($\eta$) of 2, the fraction of supernovae energy included in the winds ($\chi$) is 0.06, a maximum free travel time of 10 per cent of the Hubble time, and particles remain decoupled from hydrodynamics until the density of the particle is less than 10 per cent of the critical star formation density.

Dark matter halo substructure is inserted into these base galaxy models following a procedure similar to \cite{gauth2006}; the masses of subhaloes are sampled from the mass function from \cite{gao2004},
\begin{equation}
\dfrac{dN}{dM} = 10^{-3.2}\left(\dfrac{M_{\rm{sat}}}{h^{-1}M_{\odot}}\right)^{-1.9}  \left(\dfrac{1}{h^{-1}M_{\odot}}\right)^{2},
\end{equation} 
\noindent
and dark matter haloes are placed according to the cumulative number density in \cite{gao2004},

\begin{equation}
N = N_{\rm{total}}x^{2.75}\dfrac{(1 + 0.244x_{s})}{(1 + 0.244x_{s}x^{2})}.
\end{equation}
\noindent
Here, $N_{\rm{total}}$ is the total number of subhaloes contained within the virial radius ($r_{200}$), $x = r/r_{200}$, and $x_{s} = r_{s}/r_{200}$. Subhaloes are truncated at their tidal radius, using the Jacobi approximation from \cite{binney2008}, similar to \cite{gauth2006}. Also similar to \cite{gauth2006}, the centre-of-mass velocity of the subhaloes is set by solving the Jeans equation at the centre-of-mass of the subhalo within the potential of the parent halo, 

\begin{equation}
\dfrac{d(N\overline{\sigma_{r}^{2}})}{dr} + \dfrac{N}{r}\left[2\overline{\sigma_{r}^{2}} - \left(\overline{\sigma_{\theta}^{2}} + \overline{\sigma_{\phi}^{2}} \right) \right] = - N \dfrac{d\Phi}{dr},
\end{equation}

\noindent
where $\overline{\sigma_{r}^{2}}$, $\overline{\sigma_{\theta}^{2}}$, and $\overline{\sigma_{\phi}^{2}}$ are the mean-squared velocity dispersions in spherical coordinates, and $\Phi$ is the gravitational potential of the parent dark matter halo. The relation between the radial and tangential dispersions are determined by the anisotropy parameter, $\beta$, which is fixed at $\beta = 0.3$ similar to \cite{gauth2006}, 

\begin{equation}
    \beta = 1 - \dfrac{\sigma_{\theta}^{2} + \sigma_{\phi}^{2}}{2\sigma_{r}^{2}}.
\end{equation}

\noindent
It should be noted that the total subhalo mass function is typically dominated at the higher-mass end with larger subhaloes in our higher-mass samples containing log$_{10}$(M/$M_{\odot}$) $\gtrsim$ 10, which is relevant for considering disc heating mechanisms.

Hot gas coronae have been observed in Milky Way-like systems \citep[e.g.][]{moster2011}. Thus, these coronae are included in our host galaxies and subhaloes given the significant reservoir of gas present in these structures. We assume a hot gas component equivalent to 10 per cent of the other galactic components combined, similar to \cite{kauf2007}. For smaller-mass systems ($10^{9} - 10^{10}$ M$_{\odot}$ stellar mass), where blow-out of gas is energetically easier, this mass is adjusted to 2.5 per cent of other components. The mass density profile of the gas haloes follows the function presented in \cite{fang2020},

\begin{equation}
\rho_{\rm{G}}(r) = \dfrac{\rho_{0}}{(r + r_{1})^{\alpha_{1}} (r + r_{2})^{\alpha_{2}}}.
\end{equation}
\noindent
Following \cite{fang2020}, we use $r_{1} = 0.75 r_{s}$, $r_{2} = 100$ kpc, $\alpha_{1} = 1$, and $\alpha_{2} = 2$, where $r_{s}$ is the scale radius of the dark matter halo. For the lower-mass galaxies, we set $r_{2} = 50$ kpc to account for galaxies being roughly half the size of Milky Way-like galaxies. 

The temperature of gas particles in gas haloes at a radius $r$ is calculated based on the cumulative mass of the dark matter and gas contained within $r$ and the density of the hot gas \citep{kauf2007}, 

\begin{equation}
T(r) = \dfrac{\mu m_{p}}{k_{B}} \dfrac{1}{\rho_{\rm{G}}(r)} \int_{r}^{\infty}\rho_{\rm{G}}(r)\dfrac{GM(r)}{r^{2}} dr,
\end{equation} 
\noindent
where $\mu$, $m_{p}$, and $k_{B}$ are the mean molecular weight, the mass of a proton, and the Boltzmann constant, respectively. The corona gas is assumed to be fully ionized.

\subsection{Parameter Determination}

To gain insight on the influence that mass has on the evolution of SFR properties, we consider both low-mass ($10^{9} - 10^{10}$ M$_{\odot}$ stellar mass) and higher-mass ($10^{10} - 10^{11}$ M$_{\odot}$ stellar mass) galaxies. Modelled galaxies are constructed on the basis of observed structural parameters from the Sloan Digital Sky Survey \citep[SDSS,][]{york2000}. Specifically, disc scale lengths, bulge effective radii, and redshift estimates are taken from \cite{simard2011} and stellar mass estimates for the disc and bulge components are taken from \cite{mendel2014}.

\begin{figure*}
    \includegraphics[scale=0.4]{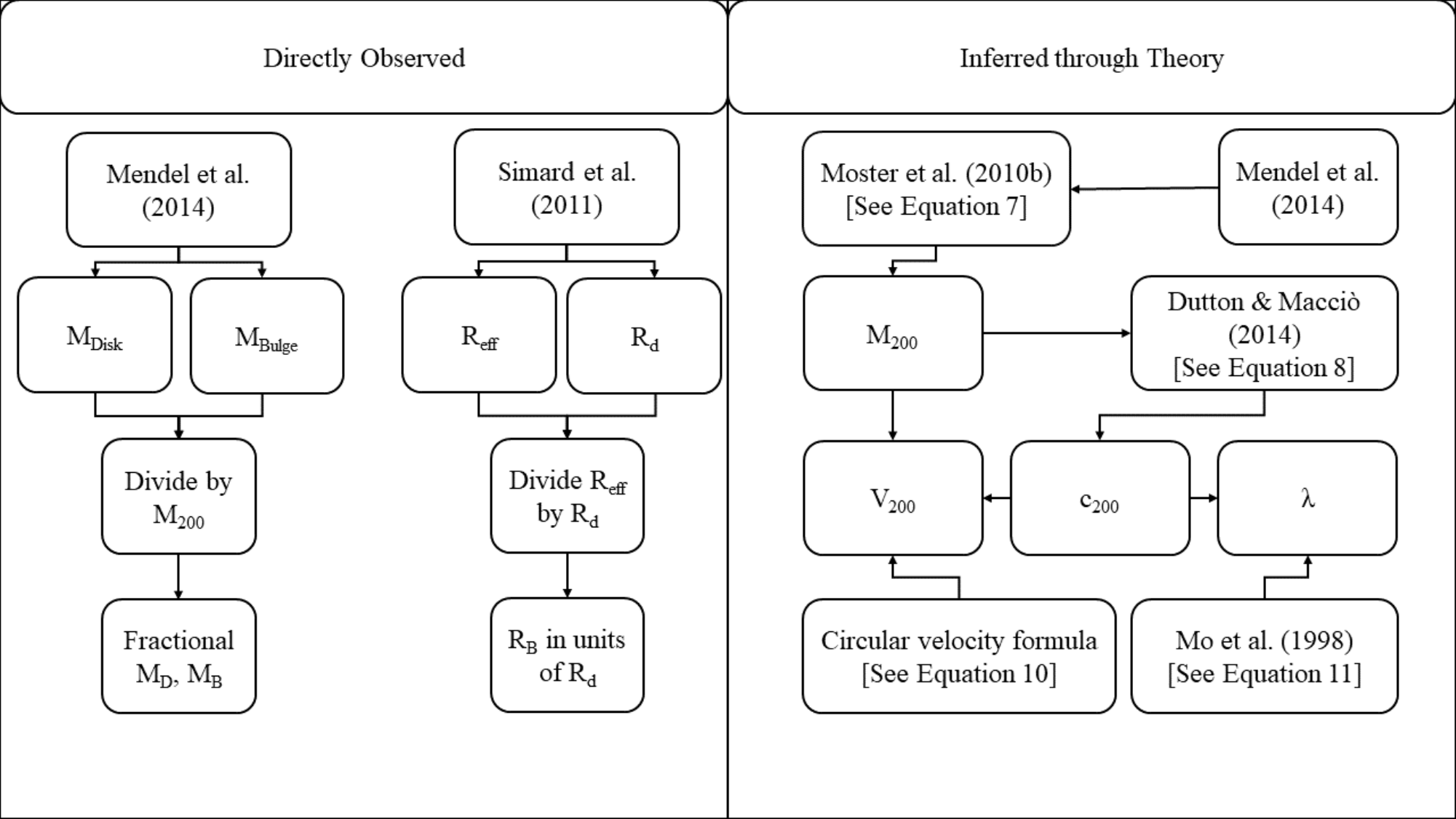}
    \caption{Flow chart describing the origin of model galaxy parameters. Note that this simplified view does not reflect the logical order of calculations.}
    \label{flow_chart}
\end{figure*}

The sample of 20 distinct galaxy models were constructed using kernel density estimation (KDE). A multivariate Gaussian kernel was assumed given the large sample size ($\sim$658,000 galaxies). A summary of the origin of the 6 MakeNewDisk parameters, whether directly through observational data or inferred through theoretical relations, is given in Figure~\ref{flow_chart}. The parameters required are halo concentration ($c_{200}$), halo circular velocity at the virial radius ($V_{c}(r_{200}) = V_{200}$), halo spin ($\lambda$), disc mass fraction of the total system mass ($M_{D}$), bulge mass fraction of the total system mass ($M_{B}$), and bulge scale length ($R_{B}$) in units of disc scale length ($R_{d}$).

The circular velocity at the virial radius of the halo can be directly related to the mass contained within the virial radius. In addition, the total halo mass can also be used to infer the concentration of the halo via the concentration-mass relation \citep[e.g.][]{dutton2014}. Hence, we first estimate the total halo mass based on the total stellar mass estimate from \cite{mendel2014} using the theoretically inferred stellar-to-halo mass relation in \cite{moster2010},

\begin{equation}
\dfrac{M_{\rm{star}}}{M_{\rm{halo}}} = 2\left(\dfrac{M_{\rm{star}}}{M_{\rm{halo}}}\right)_{0}\left[\left(\dfrac{M_{\rm{halo}}}{M_{A}}\right)^{-\delta} + \left(\dfrac{M_{\rm{halo}}}{M_{A}}\right)^{\gamma}\right]^{-1},
\end{equation}
\noindent
where $(\dfrac{M_{\rm{star}}}{M_{\rm{halo}}})_{0}$ is the normalization of the ratio, $M_{A}$ is the mass where the ratio is equal to the normalization, and $\delta$ and $\gamma$ are the parameters controlling the ratio at high and low masses, respectively. This relation is solved iteratively, using $\left(\dfrac{M_{\rm{star}}}{M_{\rm{halo}}}\right)_{0} = 0.0282$, $M_{A} = 10^{11.884}$ M$_{\odot}$, $\delta = 1.06$, and $\gamma = 0.556$ \citep[see Table 4 of][]{moster2010}. Note that \cite{moster2010} uses $\beta$ instead of $\delta$, which was replaced here to minimize confusion with the anisotropy parameter. Conventionally, we use the virial radius ($r_{200}$) as the cutoff radius of the galaxies. However, given the $r^{-4}$ nature of the Hernquist profile used, some particles, most notably dark matter, exist beyond the virial radius, with 99 per cent of mass inside 1.1$r_{200}$. Hence, we use this mass estimate as the approximate mass contained within the virial radius, denoted $M_{200}$.

Once an estimate for the halo mass is obtained, the concentration $c_{200}$ can be inferred using the concentration-mass relation. Solving the relation found in \cite{dutton2014}, we obtain an estimate for the concentration:

\begin{equation}
\rm{log}_{10}(c_{200}) = 0.905 - 0.101 \rm{log}_{10}\left(\dfrac{\textit{M}_{200}}{10^{-12}\textit{h}^{-1}}\right).
\end{equation}
\noindent
In the case of galaxies above redshift $z = 0.5$, we estimate the concentration using the approximation used in \cite{klypin2011},

\begin{equation}
c_{200}(M_{200},z) = c_{0}(z)\left(\dfrac{M_{200}}{10^{12} h^{-1} M_{\odot}}\right)^{-0.075}\left[1 + \left(\dfrac{M_{200}}{M_{0}(z)}\right)^{0.26}\right],
\end{equation}
\noindent
where $c_{0}(z)$ and $M_{0}(z)$ are provided in Table 3 of the aforementioned paper. 

Using this estimate for $c_{200}$, the virial radius is estimated by computing the halo scale radius. This is done by assuming the total halo mass can be rearranged to lie within the virial radius. Since the density within this radius is 200 times the universal critical density, the virial radius definition gives a direct relation between the mean halo density, mass, and radius. The circular velocity $V_{200}$ is then computed simply as

\begin{equation}
V_{200} = \sqrt{\dfrac{GM_{200}}{r_{200}}},
\end{equation}
\noindent
where $G$ is the universal gravitational constant.

The bulge and disc mass fractions ($M_{B}$ and $M_{D}$) are estimated as the stellar mass, from \cite{mendel2014}, of the respective component divided by the calculated $M_{200}$. The bulge scale length $R_{B}$ is approximated using the bulge effective radius and the disc scale length from \cite{simard2011} since MakeNewDisk requires that the bulge scale length is in units of the disc scale length.

The halo spin ($\lambda$) is estimated using the size-$\lambda$ relation first proposed in \cite{mo1998},

\begin{equation}
R_{d} = \frac{1}{\sqrt{2}}\left(\dfrac{j_{d}}{M_{D}}\right) r_{200} f_{c}^{-0.5} f_{R}\lambda,
\end{equation}
\noindent
where $j_{d}$ is the fraction of the total angular momentum that the disc contains, $M_{D}$ is the disc mass fraction, $R_{d}$ is the disc scale length, $f_{c}$ is a factor that depends on the concentration $c_{200}$, and $f_{R}$ is a factor to account for gravitational effects of the disc:

\begin{equation}
f_{R} = 2 \left[\int_{0}^{\infty}e^{-u}u^{2} \frac{V_{c}(R_{d}u)}{V_{200}} du \right]^{-1}.
\end{equation}
\noindent
The concentration factor, 

\begin{equation}
f_{c} = \frac{c_{200}}{2} \dfrac{1 - [1/(1 + c_{200})^{2}] - 2[\ln{(1 + c_{200})}/(1 + c_{200})]}{[c_{200}/(1 + c_{200}) - \ln{(1 + c_{200})}]^{2}}
\end{equation}
\noindent
is present due to energetic properties of the dark matter in the halo. We adopt the common assumption that the specific angular momentum of the disc is the same as its dark matter halo, which results in $j_{d}$ $\backsim$ $M_{D}$ \citep{mo1998}.

\subsection{Sample Determination} \label{Sample Determination}

The 6 required parameters for MakeNewDisk are calculated for each observed galaxy, and then a sample of 1,000,000 mock galaxies are produced based on the KDE. A total of approximately 134,000 and approximately 543,500 galaxies are within the low-mass and high-mass regimes, respectively.  Extrapolation of parameter distributions can naturally produce unphysical values, particularly with parameter values that are close to zero, such as halo spin. Refining the total sample by excluding unphysical parameter values and galaxies outside the corresponding stellar mass bin, we sample our final 10 galaxies from both the low-mass and high-mass bins, resulting in approximately 40,000 and 330,000 available galaxies in the low- and high-mass bins, respectively. This leads to rejecting about 70 per cent of generated low-mass galaxies and about 39 per cent of high-mass ones. As an additional criterion, we require that the disc and bulge mass fractions are within 10 per cent of the generating sample average (after nonphysical parameters are removed). The derived galaxy parameters are summarized in Tables~\ref{tab1} and~\ref{tab2}.

\begin{table*}
\noindent
\centering
\scalebox{1.0}{
\begin{tabular}{|c c c c c c c c c c c c c c c|}
\hline
Galaxy & \vline & $c_{200}$ & \vline & $V_{200}$ [km s$^{-1}$] & \vline & $\lambda$ & \vline & $M_{D}$ & \vline & $M_{B}$ & \vline & $R_{B}$ & \vline & Total Mass [$10^{10}$ M$_{\odot}$] \\ [0.5ex]
\hline

1 & \vline & 9.56 & \vline & 104.04 & \vline & 0.045 & \vline & 0.01355 & \vline & 0.00753 & \vline & 0.50 & \vline & 26.18 \\ [0.25ex]
2 & \vline & 10.06 & \vline & 116.78 & \vline & 0.066 & \vline & 0.01167 & \vline & 0.00255 & \vline & 0.94 & \vline & 37.03 \\ [0.25ex]
3 & \vline & 10.07 & \vline & 68.71 & \vline & 0.068 & \vline & 0.01690 & \vline & 0.00472 & \vline & 1.65 & \vline & 7.54 \\ [0.25ex]
4 & \vline & 9.51 & \vline & 128.40 & \vline & 0.036 & \vline & 0.01285 & \vline & 0.00142 & \vline & 0.26 & \vline & 49.22 \\ [0.25ex]
5& \vline & 9.97 & \vline & 122.88 & \vline & 0.045 & \vline & 0.01721 & \vline & 0.00329 & \vline & 0.67 & \vline & 43.14 \\ [0.25ex]
6 & \vline & 10.24 & \vline & 66.75 & \vline & 0.043 & \vline & 0.01814 & \vline & 0.00252 & \vline & 0.81 & \vline & 6.92 \\ [0.25ex]
7 & \vline & 9.73 & \vline & 127.97 & \vline & 0.065 & \vline & 0.01106 & \vline & 0.00803 & \vline & 1.42 & \vline & 48.73 \\ [0.25ex]
8 & \vline & 9.42 & \vline & 119.91 & \vline & 0.044 & \vline & 0.01014 & \vline & 0.01480 & \vline & 1.05 & \vline & 40.09 \\ [0.25ex]
9 & \vline & 9.72 & \vline & 129.16 & \vline & 0.027 & \vline & 0.01202 & \vline & 0.00316 & \vline & 2.45 & \vline & 50.10 \\ [0.25ex]
10 & \vline & 9.59 & \vline & 142.75 & \vline & 0.091 & \vline & 0.01328 & \vline & 0.00096 & \vline & 0.55 & \vline & 67.64 \\ [0.25ex]

\hline
\end{tabular}
}

\caption{MakeNewDisk parameters of sample galaxies in the stellar mass range $9 \leq \rm{log}_{10}(M_{*}/M_{\odot}) \leq 10$. The total masses are from the outputs of the MakeNewDisk files.}
\label{tab1}
\end{table*}

\begin{table*}
\centering

\scalebox{1.0}{
\begin{tabular}{|c c c c c c c c c c c c c c c|}

\hline
Galaxy & \vline & $c_{200}$ & \vline & $V_{200}$ [km s$^{-1}$] & \vline & $\lambda$ & \vline & $M_{D}$ & \vline & $M_{B}$ & \vline & $R_{B}$ & \vline & Total Mass [$10^{10}$ M$_{\odot}$] \\ [0.01ex]
\hline

1 & \vline & 8.32 & \vline & 225.82 & \vline & 0.037 & \vline & 0.01103 & \vline & 0.01624 & \vline & 0.90 & \vline & 267.76 \\ [0.25ex]
2 & \vline & 8.58 & \vline & 191.28 & \vline & 0.022 & \vline & 0.01723 & \vline & 0.00540 & \vline & 0.69 & \vline & 162.73 \\ [0.25ex]
3 & \vline & 7.99 & \vline & 235.71 & \vline & 0.038 & \vline & 0.01228 & \vline & 0.01646 & \vline & 0.90 & \vline & 304.50 \\ [0.25ex]
4 & \vline & 8.37 & \vline & 198.67 & \vline & 0.024 & \vline & 0.01655 & \vline & 0.01521 & \vline & 0.75 & \vline & 182.32 \\ [0.25ex]
5& \vline & 8.75 & \vline & 202.66 & \vline & 0.024 & \vline & 0.02354 & \vline & 0.00007 & \vline & 1.02 & \vline & 193.54 \\ [0.25ex]
6 & \vline & 8.07 & \vline & 252.75 & \vline & 0.021 & \vline & 0.02220 & \vline & 0.00254 & \vline & 2.29 & \vline & 375.43 \\ [0.25ex]
7 & \vline & 8.21 & \vline & 218.34 & \vline & 0.040 & \vline & 0.01674 & \vline & 0.01533 & \vline & 0.02 & \vline & 242.02 \\ [0.25ex]
8 & \vline & 8.53 & \vline & 205.37 & \vline & 0.052 & \vline & 0.01248 & \vline & 0.02051 & \vline & 0.48 & \vline & 201.40 \\ [0.25ex]
9 & \vline & 8.42 & \vline & 208.53 & \vline & 0.039 & \vline & 0.01679 & \vline & 0.01002 & \vline & 1.25 & \vline & 210.85 \\ [0.25ex]
10 & \vline & 8.30 & \vline & 216.76 & \vline & 0.031 & \vline & 0.01915 & \vline & 0.01642 & \vline & 0.82 & \vline & 236.81 \\ [0.25ex]

\hline
\end{tabular}
}

\caption{MakeNewDisk parameters of sample galaxies in the stellar mass range $10 \leq  \rm{log}_{10}(M_{*}/M_{\odot}) \leq 11$. The total masses are from the outputs of the MakeNewDisk files.}
\label{tab2}
\end{table*}

\subsection{Physical Scenario Suite and Simulation Parameters}

One of the major aims of this study is to test the impact of various physical mechanisms on the ergodicity of SFMS deviations. Thus, we simulate galaxies with various physical processes enabled: 

\begin{itemize}
    \item solely star formation
    \item star formation with substructure
    \item star formation with stellar winds enabled
    \item star formation with substructure and stellar winds enabled
\end{itemize}
\noindent
With 20 distinct galaxy models and 4 physical scenarios (every combination of having/lacking substructure and disabling/enabling stellar wind feedback), we obtain a resulting set of 80 unique simulations. 

The number of gas particles is set in MakeNewDisk such that the total number of gas particles is within 5 per cent of 10 per cent. On average, there are 1,007,156 gas particles in each low-mass galaxy and 1,003,293 in each high-mass galaxy. The number of dark matter particles is set to 2 x 10$^{5}$, the number of stellar disc particles is set to 5 x 10$^{5}$, and the number of stellar bulge particles is set to 2 x 10$^{5}$, resulting in a total of $\sim$1,900,000 particles in each model. A summary of the particle mass resolutions is given in Table~\ref{tabres}. Softening lengths are chosen to balance spatial resolution and computational resources, set to 0.1 kpc for stellar and gas particles and 0.5 kpc for dark matter particles. Halving these softening lengths produces deviations less than 0.25 per cent in SFHs. 

\begin{table*}
\centering

\scalebox{1.0}{
\begin{tabular}{|c c c c c|}

\hline
Property & \vline & $9 \leq  \rm{log}_{10}(M_{*}/M_{\odot}) \leq 10$ & \vline & $10 \leq  \rm{log}_{10}(M_{*}/M_{\odot}) \leq 11$ \\ [0.01ex]
\hline

$\overline{\rm{N}}$$_{gas}$ & \vline & 1,007,156 & \vline & 1,003,293 \\ [0.01ex]
$m_{gas,min}$ [M$_{\odot}$] & \vline & 1.3938 x 10$^{3}$ & \vline & 7.3464 x 10$^{4}$ \\ [0.01ex]
$m_{gas,max}$ [M$_{\odot}$] & \vline & 1.3018 x 10$^{4}$ & \vline & 2.8740 x 10$^{5}$ \\ [0.01ex]
$\overline{m}$$_{gas}$ [M$_{\odot}$] & \vline & 7.2777 x 10$^{3}$ & \vline & 1.6937 x 10$^{5}$ \\ [0.01ex]
\hline
$m_{DM,min}$ [M$_{\odot}$] & \vline & 3.2394 x 10$^{5}$ & \vline & 4.6121 x 10$^{6}$ \\ [0.01ex]
$m_{DM,max}$ [M$_{\odot}$] & \vline & 3.2724 x 10$^{6}$ & \vline & 1.6992 x 10$^{7}$ \\ [0.01ex]
$\overline{m}$$_{DM}$ [M$_{\odot}$] & \vline & 1.7948 x 10$^{6}$ & \vline & 1.0419 x 10$^{7}$ \\ [0.01ex]
\hline
$m_{disc,min}$ [M$_{\odot}$] & \vline & 2.2580 x 10$^{3}$ & \vline & 4.5224 x 10$^{4}$ \\ [0.01ex]
$m_{disc,max}$ [M$_{\odot}$] & \vline & 1.6168 x 10$^{4}$ & \vline & 1.5002 x 10$^{5}$ \\ [0.01ex]
$\overline{m}$$_{disc}$ [M$_{\odot}$] & \vline & 8.7494 x 10$^{3}$ & \vline & 7.2078 x 10$^{4}$ \\ [0.01ex]
\hline
$m_{bulge,min}$ [M$_{\odot}$] & \vline & 8.7133 x 10$^{2}$ & \vline & 6.7738 x 10$^{2}$ \\ [0.01ex]
$m_{bulge,max}$ [10$^{10}$ M$_{\odot}$] & \vline & 2.9666 x 10$^{4}$ & \vline & 2.5061 x 10$^{5}$ \\ [0.01ex]
$\overline{m}$$_{bulge}$ [M$_{\odot}$] & \vline & 8.8217 x 10$^{3}$ & \vline & 1.3911 x 10$^{5}$ \\ [0.01ex]

\hline
\end{tabular}
}

\caption{Average number of gas particles in our simulations along with the minimum, maximum, and average mass of each particle type in our galaxy models.}
\label{tabres}
\end{table*}

\section{Results} \label{Results}

\subsection{General Evolution}

In order to limit any effects of initial transients on our results we focus our analysis to simulation times past 700 Myr for low-mass galaxies and past 1 Gyr for high-mass galaxies. These chosen times resemble the dynamical timescales of haloes in these mass ranges, which is important to ensure the representative nature of our simulated galaxies and to limit biases that could be present at early times (such as the hot gas halo settling after initial placement). Unless otherwise stated, reported SFRs are in terms of instantaneous SFRs determined by formed stellar material within the past 5 Myr, analogous to the SFR7 parameter defined in \cite{wang2020a}.
 
\begin{figure}
	\includegraphics[width=\columnwidth]{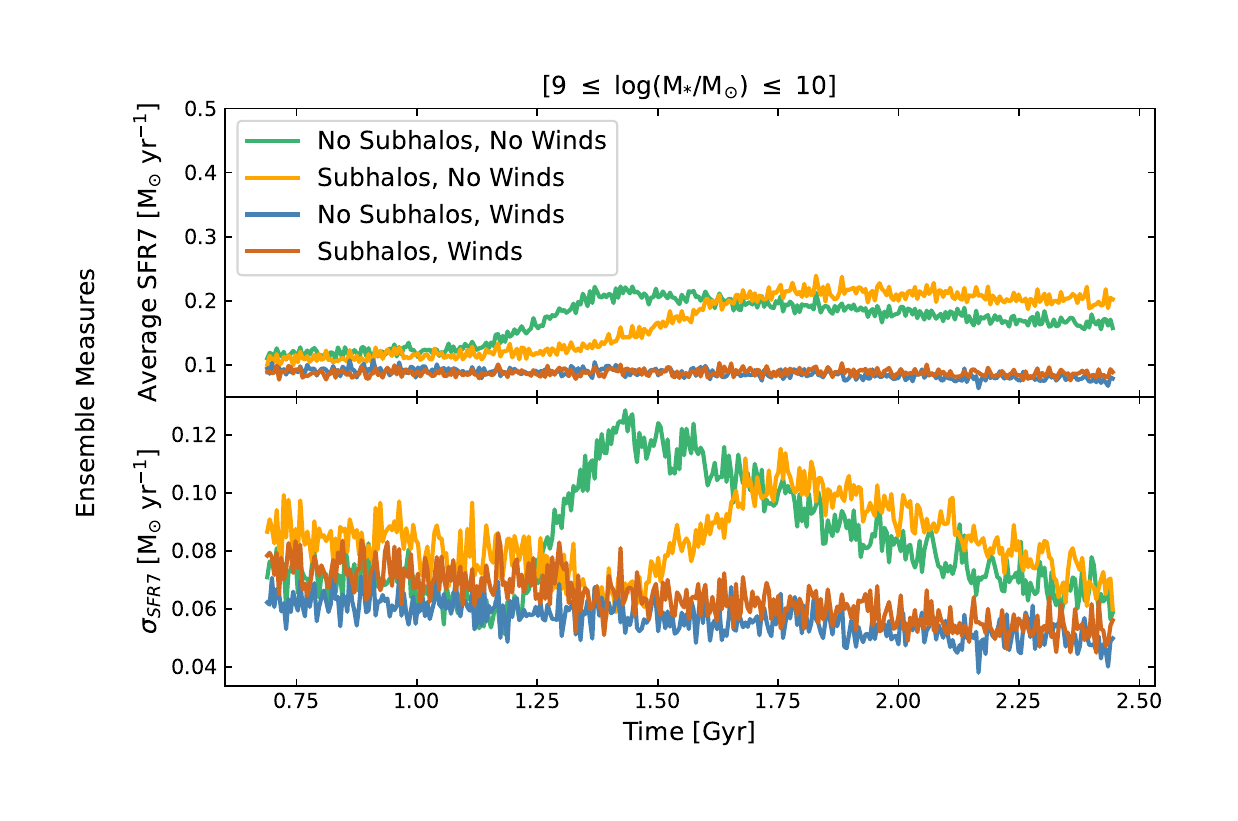}
    \caption{Ensemble averaged SFHs (top panel) and ensemble standard deviations (bottom panel) at each timestep for low-mass ($9 \leq  \rm{log}_{10}(M_{*}/M_{\odot}) \leq 10$) galaxies across the 4 investigated physical scenarios.}
    \label{sfr7_9_10}
	\includegraphics[width=\columnwidth]{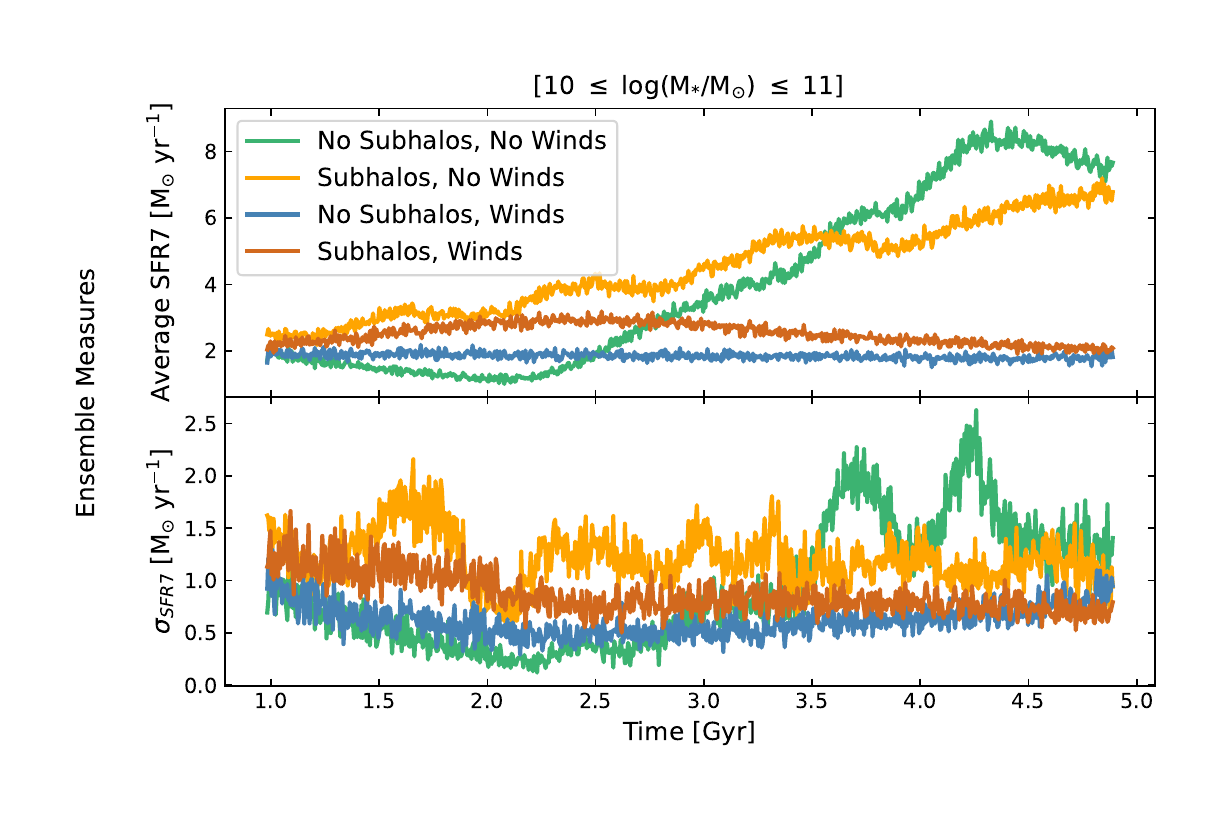}
    \caption{Ensemble averaged SFHs (top panel) and ensemble standard deviations (bottom panel) at each timestep for high-mass ($10 \leq  \rm{log}_{10}(M_{*}/M_{\odot}) \leq 11$) galaxies across the 4 investigated physical scenarios.}
    \label{sfr7_10_11}
\end{figure}

\begin{figure}
	\includegraphics[width=\columnwidth]{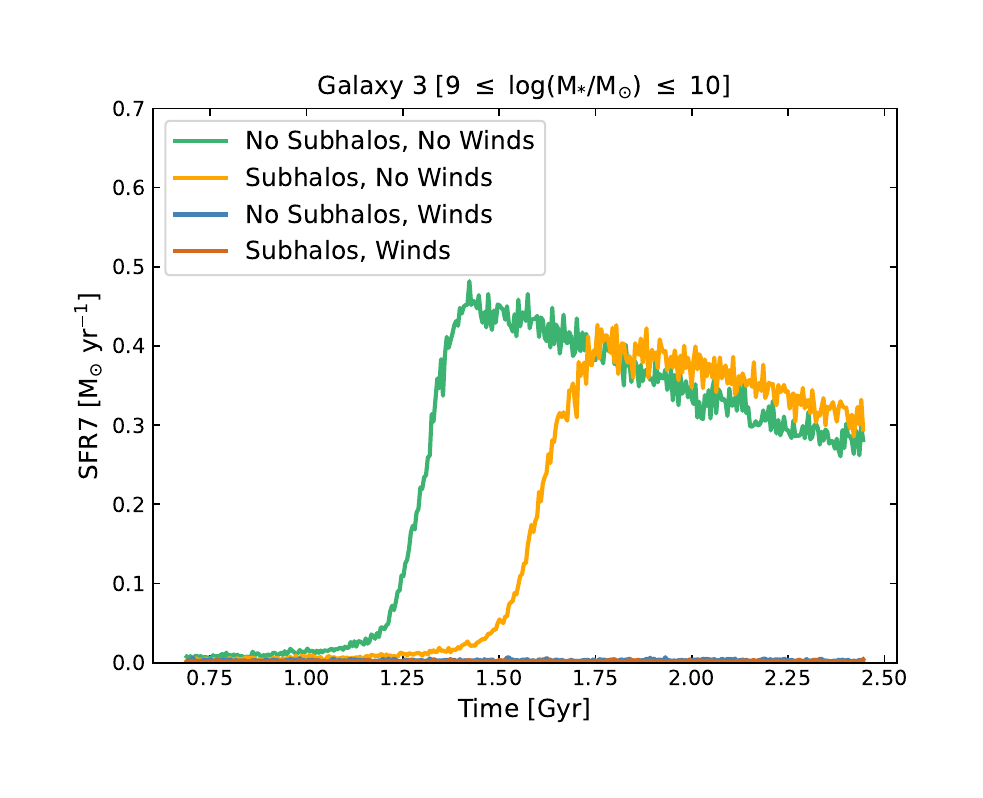}
    \caption{SFHs across the four analysed physical scenarios for galaxy model 3 in the low-mass ($9 \leq  \rm{log}_{10}(M_{*}/M_{\odot}) \leq 10$) galaxy sample.}
    \label{g3_9_10}
\end{figure}

\begin{figure}
	\includegraphics[width=\columnwidth]{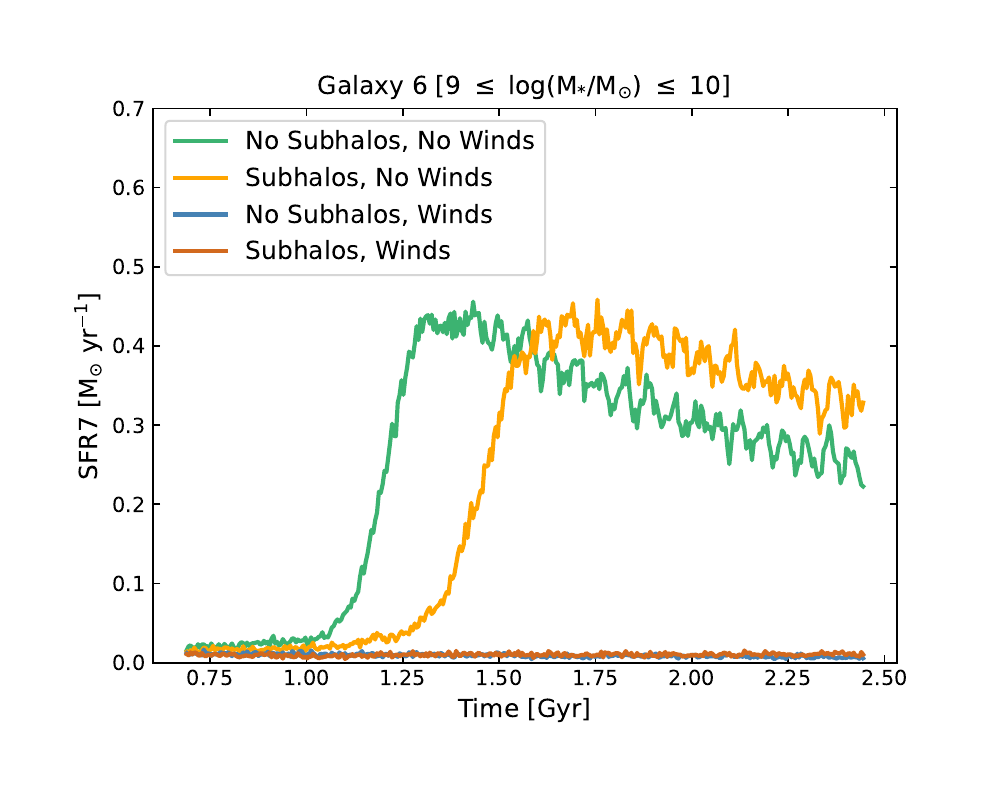}
    \caption{SFHs across the four analysed physical scenarios for galaxy model 6 in the low-mass ($9 \leq  \rm{log}_{10}(M_{*}/M_{\odot}) \leq 10$) galaxy sample.}
    \label{g6_9_10}
\end{figure}

The ensemble averaged SFHs of our low-mass sample of galaxies in each scenario are presented in Figure~\ref{sfr7_9_10}. In general, SFHs remain approximately constant until about 1.2 Gyr as star formation occurs passively in the disc, after which an increase in SFR is observed in simulations without stellar winds due to infalling gas (both from substructure and from the hot coronae). In galaxies without substructure, where the infalling gas is entirely from the corona, the increase in the average is steeper and the mean SFR is driven mainly by the two lowest mass galaxies in the sample, shown in Figures~\ref{g3_9_10} and~\ref{g6_9_10}. In runs with substructure, the increase is delayed for approximately 100 Myr. Analysis of the simulations show that this delay is caused by substructure disrupting the disc and breaking the symmetrical infall of halo gas. Past 1.75 Gyr, the mean SFR evolution of simulations with and without substructure are largely the same as all of the available gas has sufficient time to cool and infall to the disc.

In galaxy models 3 and 6 of the small-mass sample in runs without stellar winds, the higher values of the SFHs at later times are a result of having shorter cooling times and shallow potentials. The halo gas infalls to the disc similar to other galaxies, however the build-up of star-forming gas results in weak gas outflow due to the stellar feedback alone. This behaviour manifests in the SFR as a rapid rise in star formation in the disc followed by a slow decay. These two lowest-mass galaxies are further from the mass transition present in galaxies that can more easily support smooth as opposed to bursty SFHs \citep[e.g.][]{hop2023}. Conversely, for the runs with stellar winds the overall behaviour mirrors the broad behaviour seen in other runs, namely a smooth
SFH. 

The time evolutions of the ensemble averages and ensemble standard deviations for high-mass galaxies are displayed in Figure~\ref{sfr7_10_11}. Similar to the low-mass sample, SFRs of galaxies remain mostly smooth until approximately 2.0 Gyr. The average of runs with no additional physics increases more notably past 2.0 Gyr, with stellar feedback occasionally slowing this increase, until reaching a maximum at about 4.25 Gyr. Across the simulation time, the average SFR gradually increases in galaxies with substructure (and stellar winds disabled), as more close encounters with subhaloes results in more cool accretion of gas from substructure onto the disc. However, the spread of SFHs remains mostly constant, particularly past 2.5 Gyr. Simulations with stellar winds enabled remain mostly unchanged over the simulation time, as the winds serve to delay and regulate accretion both from the hot corona and from substructure. However, there is a distinction between runs with substructure and those without at early simulation times since stellar winds have a delayed impact in higher-mass/velocity systems and as star formation occurs in substructure.

In contrast with the low-mass sample, we find that simulations with no substructure (both with and without stellar winds) are more similar to each other than runs with substructure, until approximately 3.0 Gyr. This makes intuitive sense since the effect of stellar feedback and winds is not as pronounced in higher-mass systems. In addition, we find that substructure in high-mass galaxies results in higher SFRs than runs without, until approximately 3.5 Gyr after which the runs without substructure (and stellar winds disabled) produce larger SFRs. The delayed accretion and cooling of hot halo gas produces these larger SFRs, in the absence of disc thickening/heating. The opposite trend is observed in the low-mass sample, where runs with substructure exceed those without past 1.75 Gyr. We understand this as being due to less mass being contained in the hot gas coronae of the lower-mass systems, resulting in less available fuel for star formation. Stellar winds slow accretion of hot gas onto the star-forming disc, however, this effect is delayed in higher-mass systems as evidenced by the disparity between runs with stellar winds enabled (and the lack thereof in low-mass systems). This is also expected, since the higher speeds present in the higher-mass systems are less affected by the stellar winds, which have the same speeds in both high- and low-mass simulations.

We note in passing that the subgrid feedback model \citep{springel2003} used in this study has been found to produce smoother SFHs that are less able to reproduce short period variations, such as starbursts \citep[e.g.][]{sparre2017,bassini2020}. Since the time variation of the SFH is the central factor in measures of ergodicity for a single system relative to the ensemble, star formation models with distinctly different time variation behaviours are almost certain to produce different behaviours within ergodicity measures. In the most extreme example, SFHs that move smoothly along a single trend cannot, by definition, reproduce ergodic behaviour. If such models can be detrended to a constant, then while they are wide-sense stationary, their lack of time variance means they cannot be ergodic. Consequently, star formation models with more rapid and larger variations around the mean, especially those with explicitly resolved feedback such as those used in the FIRE simulations \citep[as discussed in][]{sparre2017}, likely have a higher probability of showing ergodic behaviour. However, such issues need to be set in the context of variations in SFHs that are related to changes in population behaviours, such as the trend for smoother SFHs at lower redshifts that is likely associated with more massive galaxies \citep[e.g.][]{hop2023}. For this initial study, which is constrained by lower redshift galaxy measures, using a well-studied feedback algorithm provides a useful analysis for the lower redshift population, but further research, especially on the high redshift galaxy population is clearly warranted.

\subsection{The Effect of Substructure}\label{substructure}

By visual inspection of the average SFHs, we see that substructure does not significantly impact any given galaxy's total SFR over time, despite evident disc heating, disc thickening, and axis of rotation shifts caused by close-encounters with infalling satellites. Thus, the small mass component that substructure contributes (5 per cent of the total galaxy mass) may produce clear morphological changes, but the overall star formation that occurs is not significantly affected. 

\begin{figure}
	\includegraphics[width=\columnwidth]{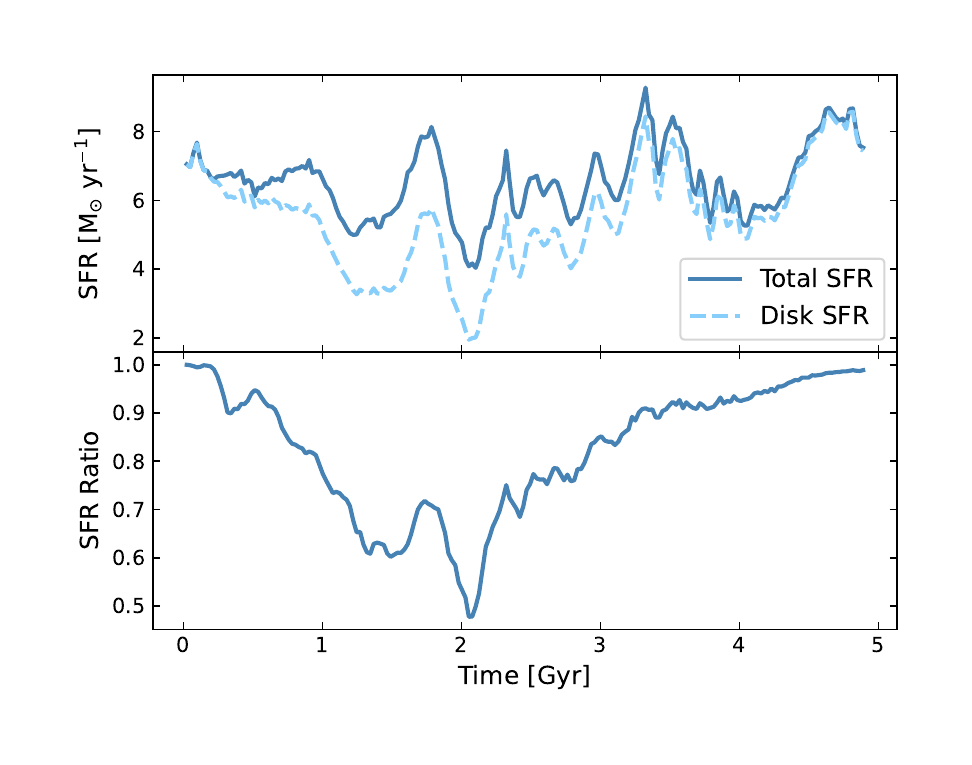}
    \caption{SFH and disc analysis for galaxy model 6 of the high-mass sample. The top panel shows the SFH of the galaxy (solid line) in addition to the SFH traced by the mock star-forming disc (dashed line). In the bottom panel, the evolution of the ratio of the disc SFR and the total SFR is shown.}
    \label{g6_sfr_disc}
\end{figure}

Quantitatively, we estimate the star formation occurring in the star-forming disc by evaluating the SFR contained within a sphere of radius 6R$_{d}$ centred at the center-of-mass of the galaxy. A sphere was found to better capture the disc SFRs without needing to adjust the disc orientation, which can change due to infalling gas. The mean fraction of SFR captured within the mock disc is 99.27 per cent over the simulated time in low-mass galaxies and 90.13 per cent of the SFR over time in high-mass galaxies. The maximum missing fraction of star formation occurring outside of the mock disc is 0.5225, which was occurs for a brief time interval in the 6th high-mass galaxy model. In the galaxy model with the largest missing fraction at any given timestep, the overall time-average of the star formation occurring within the disc is 82.67 per cent. In addition, the overall shape of the SFH is generally captured by the disc, as can be seen in Figure~\ref{g6_sfr_disc} for this high-mass galaxy.

\subsubsection{Impact of Varying Initital Subhalo Parameters}

\begin{figure}
	\includegraphics[width=\columnwidth]{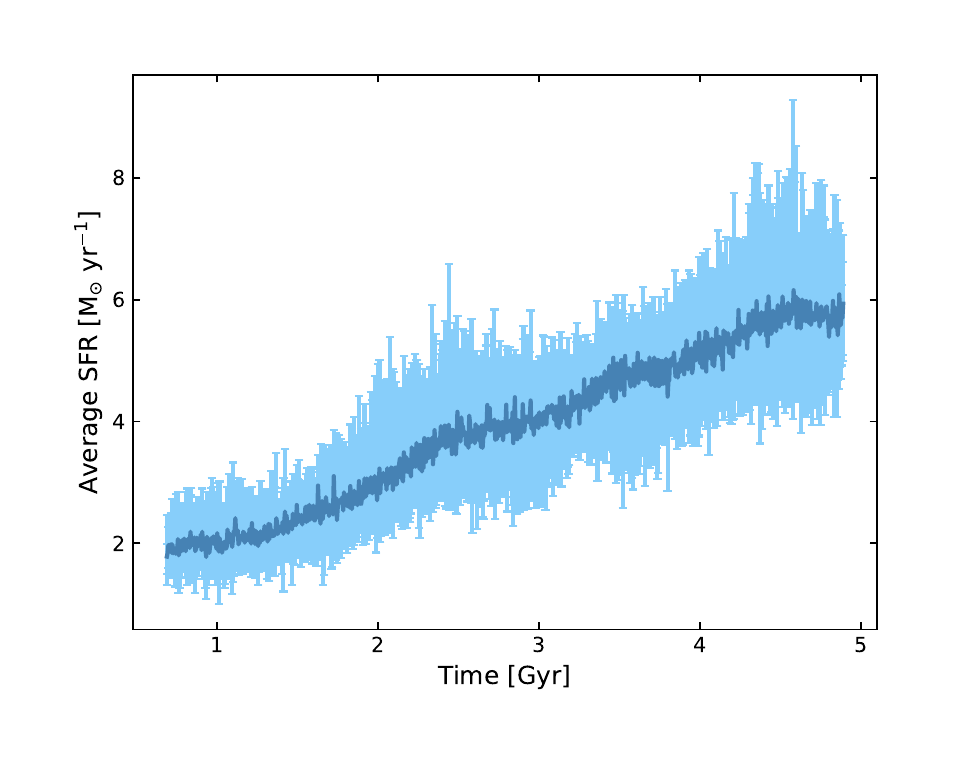}
    \caption{Average SFH of galaxy model 9 of the high-mass sample, rerun 10 times with varying initial subhalo parameters. The error bars denote the maximum and minimum deviations of the entire ensemble at each timestep.}
    \label{sub_g9_rerun}
\end{figure}

To determine the impact that the initial positions/orbits and exact mass distributions of substructure could introduce to our results, a high-mass Milky Way-like galaxy (Model 9) was simulated 10 times with varying initial subhalo parameters. The mean and range of SFHs is shown in Figure~\ref{sub_g9_rerun}. The mean deviation from the average SFH is less than 30 per cent in all simulations and there are no clear temporal dependencies on any deviations. Averaging the deviations over the simulation time, the 1-$\sigma$ deviation is 14.76 per cent of the ensemble mean ($\sigma$ = 0.5551 M$_{\odot}$ yr$^{-1}$). The largest deviation at any single timestep is within 73.41 per cent of the ensemble average, and the overall behaviour over time is consistent between runs. Thus, we find that any given initial construction of the halo substructure does not significantly impact our results.

\subsection{Tests of Ergodicity}

Given the non-stationarity of galaxy SFHs (i.e. gas consumption means that there is normally a trend), we do not expect SFRs to be ergodic. However, the deviations from the sample average (i.e. vertical deviations from the SFMS) can, in principle, be ergodic. That is, if individual galaxies differ from the SFMS in a regulated manner over time, we could derive their star formation properties based on said deviations and the well-known SFMS relation. As discussed in \cite{wang2020b}, investigating the deviations from the SFMS on different timescales produces crucial information on the variability of the SFMS on short and long timescales, in addition to what physical processes contribute most significantly to these deviations (if any). If galaxies exhibit ergodic properties, they should theoretically explore the same parameter space over time, regardless of mass. Otherwise, a galaxy's initial mass would determine its trajectory in the parameter space, breaking ergodicity. Perhaps unsurprisingly, this mass independence is suggestive of considering specific (i.e. mass normalized) quantities. 

\begin{figure}
	\includegraphics[width=\columnwidth]{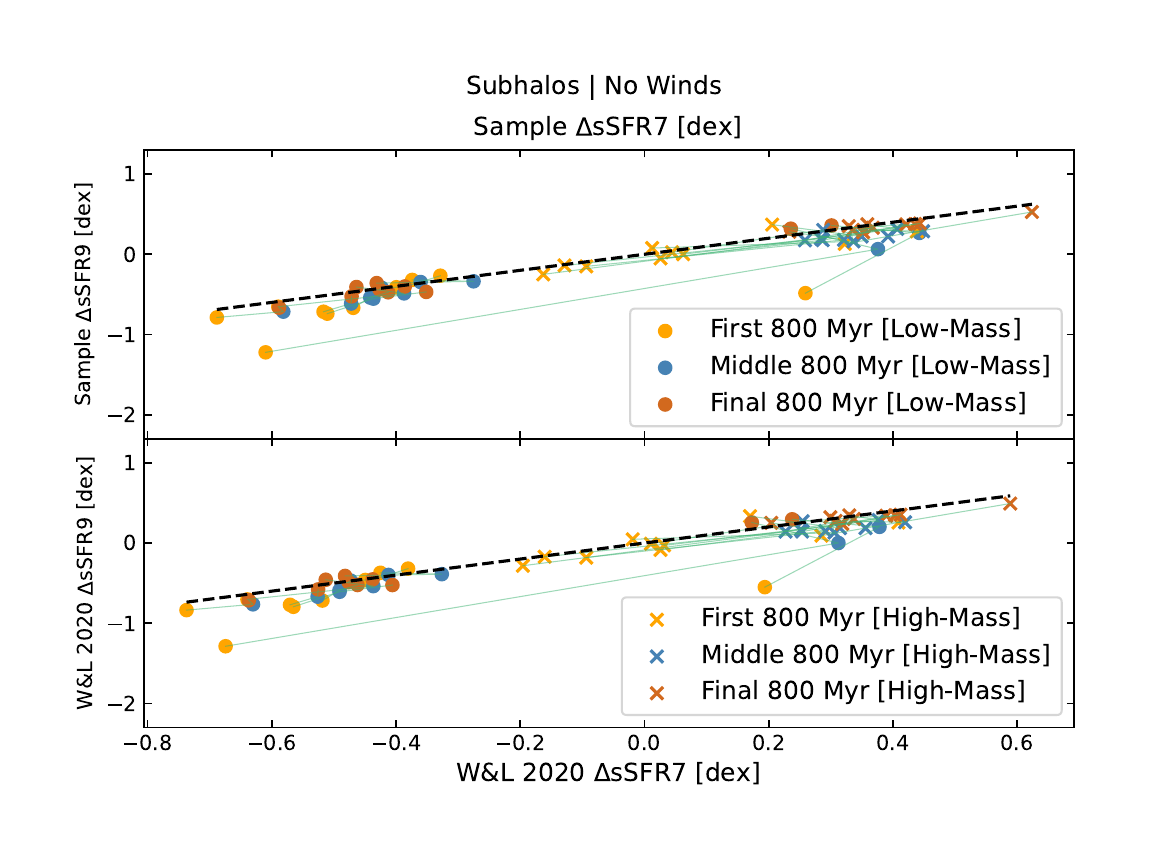}
    \caption{Relation between SFMS deviations on long timescales and short timescales (SFR9 and and SFR7, respectively) for low-mass and high-mass galaxies in our sample in simulations with substructure (and stellar winds disabled). In the top panel, the vertical deviations are calculated with respect to the sample SFMS. In the bottom panel, the deviations are calculated with respect to the SFMSs found in \protect\cite{wang2020b}. For each galaxy, deviations are calculated at 3 epochs, with low-mass galaxies denoted using points and high-mass galaxies with crosses. The dashed line indicates equality between the parameters.}
    \label{ssfr9_ssfr7_sub}
\end{figure}

In Figure~\ref{ssfr9_ssfr7_sub}, we show an example of the correlations between deviations from the SFMS on short and long timescales as measured on 5 Myr and 800 Myr timescales using the SFR7 and SFR9 parameters defined in \cite{wang2020a}. In the top panel, vertical deviations from the SFR7 and SFR9 SFMSs defined by our sample are shown. To construct the SFMS, we follow a method similar to \cite{belfiore2018}, consistent with \cite{wang2019} and \cite{wang2020b}; we divide our stellar mass range into 4 equally-spaced (logarithmically) bins and take the median SFR7 in each bin, considered over the full simulation time. We define the SFMS as the best-fit line to these points. In the bottom panel, our simulated galaxies are compared to the SFR7 and SFR9 SFMSs presented in \cite{wang2019} and \cite{wang2020b}. There is a separation between high- and low-mass galaxies, where low-mass galaxies are below the SFMS (negative deviation values) compared to high-mass galaxies occupying regions of the parameter space closer to zero. Due to the separation between low-mass and high-mass galaxies in the $\Delta$sSFR7-$\Delta$sSFR9 parameter space present in our sample, we interpret the simulated galaxies as showing signs of weak ergodicity breaking.

\begin{figure}
	\includegraphics[width=\columnwidth]{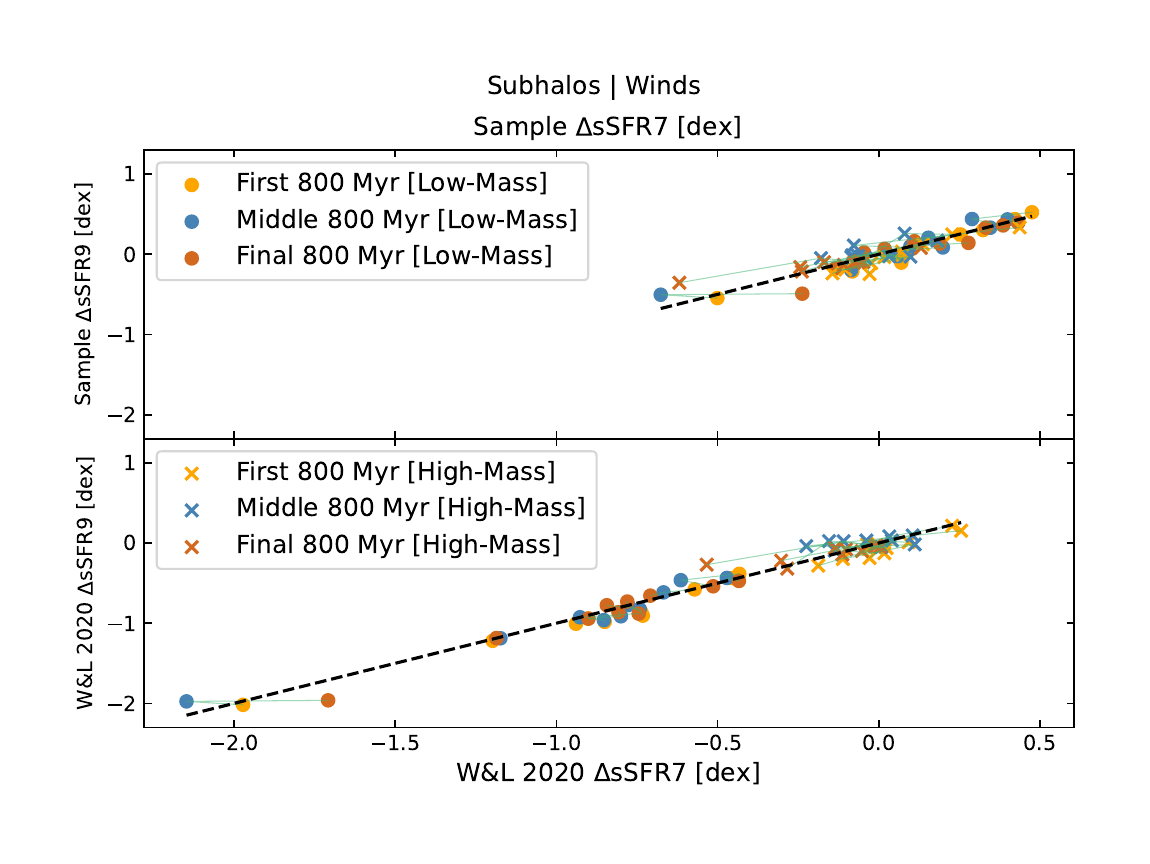}
    \caption{Similar to Figure~\protect\ref{ssfr9_ssfr7_sub}, except for simulations with substructure and stellar winds enabled. In the top panel, the vertical deviations are calculated with respect to the sample SFMS. In the bottom panel, the deviations are calculated with respect to the SFMSs found in \protect\cite{wang2020b}. For each galaxy, deviations are calculated at 3 epochs, with low-mass galaxies denoted using points and high-mass galaxies with crosses. The dashed line indicates equality between the parameters.}
    \label{ssfr9_ssfr7_wind}
\end{figure}

This separation is present across all scenarios when comparing to the SFMS in \cite{wang2019}, however when comparing to calculated sample SFMSs there are two exceptions, namely the separation is not evident for simulations with stellar winds (i.e. with and without substructure). We show the run with subhaloes and stellar winds in Figure~\ref{ssfr9_ssfr7_wind}. The resulting SFMSs constructed using galaxies with winds enabled are steeper than \cite{wang2019}, with a slope of 1.48-1.58 in runs without and with substructure, respectively, compared to 0.71. Hence, weak ergodicity breaking is not apparent across all physical scenarios analysed here. Consequently, in order to further scrutinize the apparent absence of ergodicity, we also measure ergodicity quantitatively.

To quantify the disparity from true ergodicity, we use the Thirumalai-Mountain (TM) metric \citep{thirumalai1989}, defined as the ensemble mean-squared deviation of each galaxy's time-average from the ensemble-averaged time-average,
\begin{equation}
\Omega_{e}(t) = \frac{1}{N} \sum^{N}_{i=1} [\epsilon_{i}(t) - \overline{\epsilon}(t)]^{2},
\end{equation}
\noindent
where $\epsilon(t)$ is the time average of a process over the interval $t$, $\overline{\epsilon}(t)$ is the ensemble average of these time averages, and $N$ is the size of the ensemble. 

\begin{figure}
	\includegraphics[width=\columnwidth]{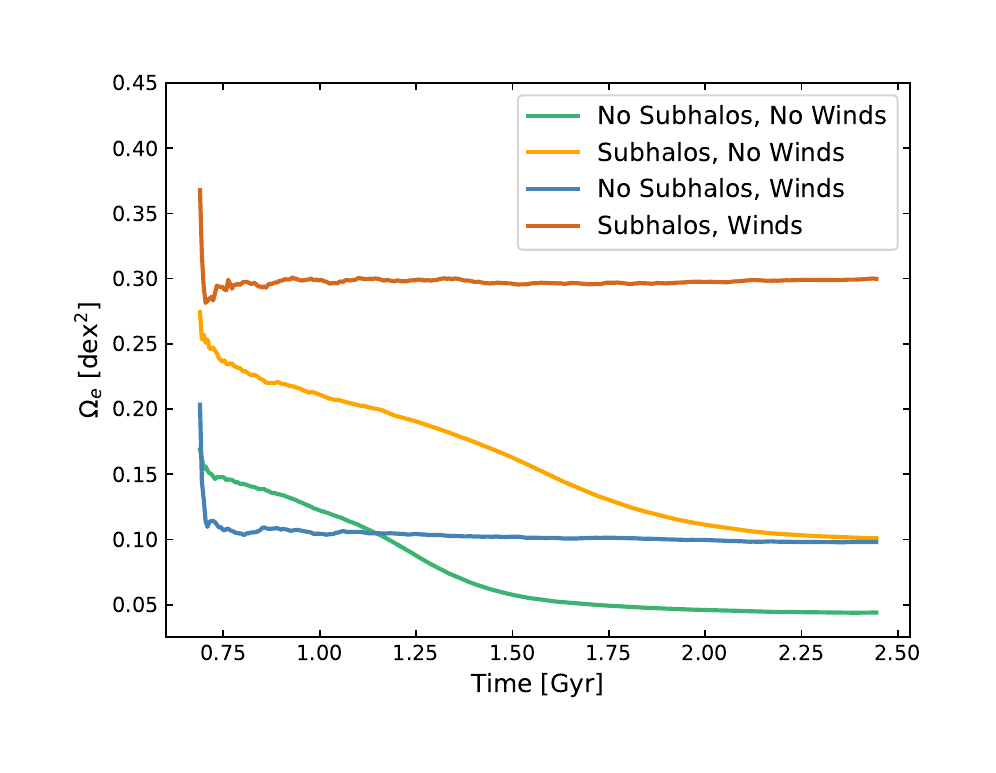}
    \caption{TM metric for SFMS deviations over the analysed time for all galaxies in our sample. In the unlikely event that the sSFR is zero at a given timestep, which occurs at 3 timesteps for a single simulation with substructure and stellar winds enabled, the timestep is ignored to avoid taking the logarithm of zero.}
    \label{tm_metric}
\end{figure}

This metric approaches zero for ergodic systems as individuals explore the full parameter space over time. In Figure~\ref{tm_metric}, we calculate the TM metric over time for vertical deviations in log-space from the \cite{wang2019} SFMS across our physical scenarios. Across all scenarios, the TM metric reaches a minimum and does not approach zero with time, even in the case where the sSFHs of galaxies with stellar winds enabled have apparently constant (stationary) ensemble averages. In the simulations with stellar winds, we find that the TM metric is mostly constant with time due to the near constant SFHs resulting from suppressed star formation and inflow of halo/subhalo gas. In runs without winds, the TM metric does significantly decrease with time until about 1.75 Gyr for runs without substructure and about 2.0 Gyr for runs with substructure. This is a consequence of different halo/subhalo gas infall times for each galaxy. Afterwards, the TM metric similarly approaches a constant value, as all galaxies reach a similar stage of evolution where the most significant infall has reached the star-forming disc and the initial burst of star formation has occurred. Thus, no scenario approaches true ergodicity in the time analysed despite the promising similarities between SFHs when stellar winds are enabled. We do note, however, that runs with stellar winds enabled lack sufficient variability to match the width of the SFMS in \cite{wang2019}. Hence, while runs with winds may appear initially to have ergodic properties in that the SFHs are stationary, these simulations are not representative of galaxies in general. 

\subsection{Partial Ergodicity}

Given the nature of how the SFH of a galaxy does not perfectly follow the SFMS trend over its evolution (i.e. the mean SFR is not stationary), it is unsurprising that galaxies are not strictly ergodic. Thus, we find that there is ergodicity breaking in the sSFR deviations from the SFMS. Specifically, since there is no physical/theoretical barrier preventing low-mass and high-mass galaxies from occupying the same areas of the normalized SFMS deviation parameter space (in the form of vertical sSFRs), this is a case of weak ergodicity breaking (as opposed to strong breaking). Intuitively, by averaging over the (apparently, although not strictly) disjoint sets in the parameter space, it is possible to construct an average pseudo-trajectory that explores the full parameter space (e.g. consider constructing average particle trajectories from distinct sub-volumes). If by averaging over low-mass and high-mass sSFRs over time we recover a time-averaged sSFH that resembles the ensemble-average at any given time, we would classify this behaviour as 'partially ergodic'; although it is not truly ergodic in the traditional sense, the constructed system displays ergodic properties. As an intuitive example, consider a container of gas and liquid. Within both phases the system could be considered ergodic, however the overall system is not. Such a system can then be considered partially ergodic, with the ergodicity restricted to subspaces. This concept has been considered in other fields, particularly in mathematical contexts \citep[see the ergodic decomposition theorem,][]{gray2009}.

\begin{figure}
	\includegraphics[width=\columnwidth]{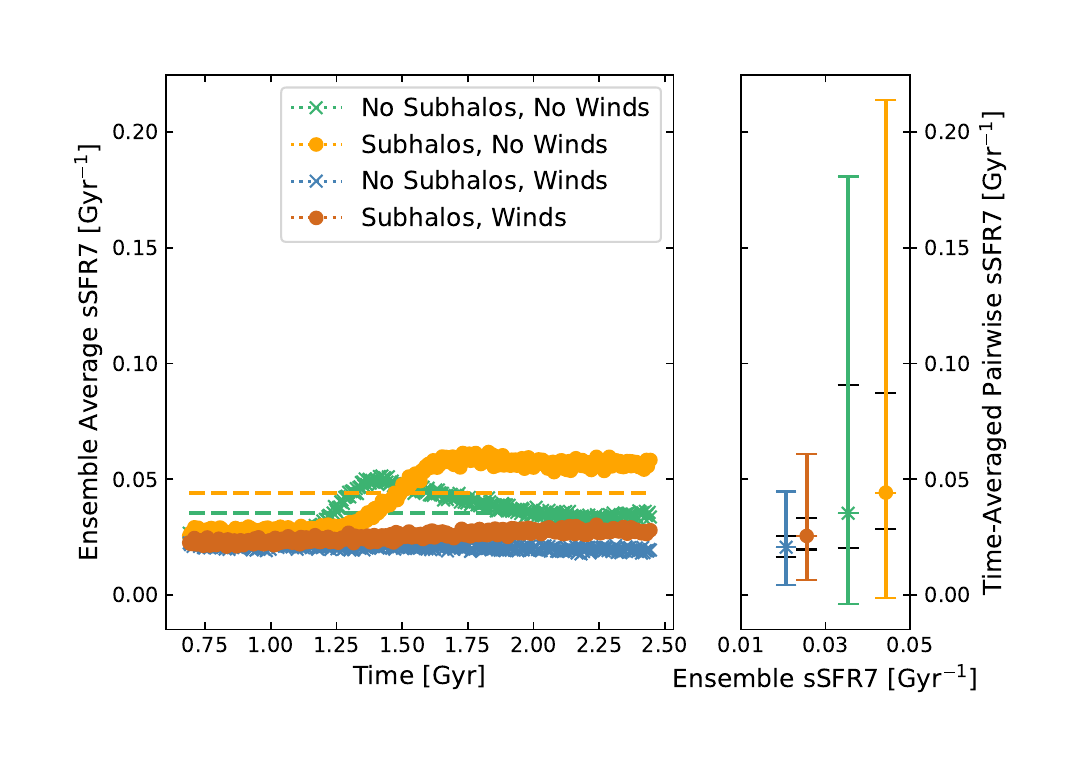}
    \caption{Test of partial ergodicity of sSFRs in our sample. The total ensemble average sSFRs at each timestep are displayed in the left panel. The time average of each scenario is represented using dashed-lines. In the right panel, the ranges of deviations for pairwise sSFHs are shown. The coloured error bars depict the minimum and maximum deviations and the black error bars are the average deviations across all timesteps.}
    \label{demi_ergodic}
\end{figure}

In Figure~\ref{demi_ergodic}, we test this theory by averaging the sSFRs of pairs of galaxies (one low-mass, the other high-mass). At each timestep, all possible pairs of low- and high-mass galaxies are averaged over their sSFRs. The maximum and minimum deviations of the pairwise-averaged sSFRs from the time-averaged ensemble averages are calculated at each timestep. The maximum deviations and the minimum deviations of all timesteps are displayed with coloured error bars and the average deviations across the simulation time are displayed with black error bars. The maximal deviation for these pairwise-averaged sSFHs are within 0.61 dex, 0.58 dex, 0.07 dex, and 0.14 dex of the ensemble mean for runs with no additional physics, no stellar winds, stellar winds and no substructure, and substructure and stellar winds, respectively. Given the varying ensemble-average over time, the larger deviations present in simulations without stellar winds is unsurprising, especially considering the notable deviation between the time-average and the ensemble-average past 1.75 Gyr and at $\sim$1.5 Gyr for runs with and without substructure, respectively. Despite the larger range of pairwise sSFR7 values, the time-average of deviations greater than the ensemble-average over all timesteps are 0.20 dex, -0.01 dex, -0.63 dex, and -0.52 dex, respectively. Similarly, the time-average of deviations below the ensemble-average are -0.37 dex, -0.45 dex, -0.71 dex, and -0.64 dex, respectively. Thus, the time averages of sSFHs constructed by averaging over galactic mass bins resemble the ensemble averages at any given time, well within the accepted bounds of the SFMS \citep[0.2 - 0.3 dex, e.g.][]{brinchmann2000,scholz2023}. This is indicative that our constructed sSFHs are at least partially ergodic; while not truly ergodic, sSFHs constructed in this manner exhibit ergodic properties.

\section{Conclusions} \label{Conclusions}

We have tested the amount of ergodicity of star formation properties in galaxies and an initial test into the physical mechanisms responsible for ergodic deviations. Our main conclusions are the following: 
\begin{itemize}
    \item A clear separation between SFMS deviations of low-mass and high-mass galaxies exists for the simulation sample, demonstrating a weak breaking of ergodicity. In addition, in no physical scenario does the TM metric approach a value of zero. Thus, in general, simulated galaxies are not ergodic.

    \item Evidence for partial ergodicity is found, where averaging over pairs of sSFHs in separate mass bins result in time-averaged sSFHs resembling the ensemble average sSFH at any given time. Since this is not true ergodicity, we refer to this property as 'partial' ergodicity.

    \item Substructure does not significantly alter the SFHs of galaxies, however larger subhaloes in high-mass systems do cause notable disc thickening and heating from close encounters, in agreement with \cite{grand2016}. 

\end{itemize}

Despite nearly stationary average sSFHs, simulations with stellar winds lack sufficient variability to be considered truly ergodic. The subtleties involved in defining the bounds of ergodicity are not obvious; while stationarity is necessary for ergodicity, it is not a sufficient condition as evidenced by the TM metric analysis (Figure~\ref{tm_metric}). Thus, while we can restrict simulated galaxies to tight confines of the parameter space in order for individual averages to resemble the population averages, this is neither realistic nor useful for inferring individual galaxy evolution from population trends for standard observed systems. 

Future work will include the effects of mergers and environments, which is beyond the scope of this initial study; the motivation for this work was to constrain the ergodicity of SFRs in galaxies and how different physical processes affected ergodicity. The addition of environment and major mergers introduces another confounding variable on ergodicity, which contrasts with the aims of this study. A study into ergodicity using populations drawn from cosmological simulations, especially looking at high redshift evolution, could potentially add significant further insight into this research area. While there may be some challenges with reproducing precise structural statistics, for example disc and bulge sizes, relative to observed systems, such models provide key insight into the role of overall system mass as well as mergers and mass accretion. Additionally, different feedback models may well have statistically relevant impacts alongside the impact of additional feedback processes being considered, such as active galactic nuclei. We plan to look at these additional important issues in future research. 

\section*{Acknowledgements}

We thank the anonymous Reviewer for their insightful comments that improved the paper. We would like to thank Dr. Volker Springel for providing access to the MakeNewDisk package. Funding from National Sciences and Engineering Research Council of Canada, Canada Research Chairs Programme, and Research Nova Scotia are acknowledged. Simulations were run on the Saint Mary's Computational Astrophysics Laboratory. FMS acknowledges support from the Saint Mary's Faculty of Graduate Studies and Research Graduate Awards, the Father Burke Gaffney Memorial Scholarship, and the John Despard de Blois Scholarship for his Masters degree during which the bulk of this research was completed.

The analysis presented in this work made use of the \textbf{python} packages \textbf{NumPy} \citep{harris2020}, \textbf{SciPy} \citep{scipy}, and \textbf{Matplotlib} \citep{hunter2007}.

\section*{Data Availability}

The galaxy parameter data analysed in the article are available in the Centre de Donn\'ees astronomiques de Strasbourg (CDS), at \url{https://cdsarc.cds.unistra.fr/viz-bin/cat/J/ApJS/196/11} [10.26093/cds/vizier.21960011] and \url{https://cdsarc.cds.unistra.fr/viz-bin/cat/J/ApJS/210/3} [10.26093/cds/vizier.22100003].



\bibliographystyle{mnras}
\bibliography{ref} 






\bsp	
\label{lastpage}
\end{document}